\documentclass[sigconf,screen]{acmart}

\usepackage{enumitem}
\usepackage{tabularx}
\usepackage{xurl}  
\usepackage{graphicx}
\usepackage{amsmath}
\usepackage{multirow, multicol, makecell}
\usepackage{array}
\usepackage{url}
\usepackage{textcomp}
\usepackage{stfloats}
\usepackage{xcolor}
\usepackage{xspace}
\usepackage{booktabs}
\usepackage{framed}
\usepackage{mdframed}
\usepackage{colortbl}
\usepackage{hyperref}
\usepackage[normalem]{ulem}  % for \sout
\usepackage{microtype}

\newcommand{\added}[1]{#1}

\newcommand{\removed}[1]{}

\definecolor{boxbg}{rgb}{0.9,1,0.9} % Light green background
\definecolor{boxborder}{rgb}{0.0,0.3,0.1} % Dark green border
\definecolor{darkgreen}{rgb}{0.0, 0.5, 0.0}

\newcommand{\totalCount}{124,757\xspace}
\newcommand{\totaltoxic}{946\xspace}

\newcounter{observationcomment@counter}
\title{A Low-Cost Human-in-the-Loop Investigation of Toxicity on GitHub at Scale}
\author{Rahat Rizvi Rahman}
\orcid{0009-0006-7143-2355}
\affiliation{%
  \institution{Virginia Commonwealth University}
  \city{Richmond}
  \country{USA}
}
\email{rahmanr12@vcu.edu}

\author{Mia Mohammad Imran}
\orcid{0000-0003-2477-9971}
\affiliation{%
  \institution{Missouri University of Science and Technology}
  \city{Rolla}
  \country{USA}
}
\email{imranm@mst.edu}

\author{Kostadin Damevski}
\orcid{0000-0001-7799-2026}
\affiliation{%
  \institution{Virginia Commonwealth University}
  \city{Richmond}
  \country{USA}
}
\email{kdamevski@vcu.edu}

\copyrightyear{2026}
\acmYear{2026}
\setcopyright{cc}
\setcctype{by}
\acmConference[ASE '26]{Proceedings of the 41st IEEE/ACM International Conference on Automated Software Engineering}{October 12--16, 2026}{Munich, Germany}
\acmBooktitle{Proceedings of the 41st IEEE/ACM International Conference on Automated Software Engineering (ASE '26), October 12--16, 2026, Munich, Germany}
\acmDOI{10.1145/3832783.3837505}
\acmISBN{979-8-4007-2882-2/2026/10}
\acmSubmissionID{ase26main-p1633-p}
\received{2026-03-26}
\received[accepted]{2026-06-18}

\begin{document}




\sloppy

\begin{abstract}
Toxic interactions in open source discussions can alienate contributors and threaten project sustainability, yet prior empirical studies of GitHub toxicity have been limited in scale, raising questions about their generalizability. Scaling up is difficult because toxicity on GitHub is often implicit and context-dependent, making both fully manual annotation and LLM-based labeling unreliable.

We present a human-in-the-loop (HITL) annotation methodology that makes large-scale, domain-calibrated toxicity labeling practical. A single call to a small, local LLM produces both a toxicity prediction and a set of interpretable event category scores. A lightweight Random Forest validator then uses those scores to flag the small subset of conversations most likely to be mislabeled, directing human review only where it is needed. The validator outperforms confidence-based and multi-LLM baselines while adding low annotation cost.

We apply this pipeline to over 124,000 GitHub issue and pull request conversations. Using the resulting dataset, we evaluate key findings from prior small-scale research, confirming some and qualifying others, and present new insights into the prevalence, characteristics, and dynamics of toxicity across diverse open source projects.

\end{abstract}

\maketitle

\section{Introduction}

Open source collaboration on GitHub has enabled substantial innovation, but it has also created space for toxic discourse in project discussions. Recent research suggests that toxicity on GitHub differs in important ways from toxicity on other online platforms~\cite{miller_did_2022,schluger_proactive_2022,ferreira2021shut}. Rather than appearing primarily as overt hostility, it often takes subtler, more technical forms, such as dismissive language, passive-aggressive comments, condescending code reviews, and exclusionary feedback toward newcomers. These behaviors can discourage participation, especially among less experienced developers, which in turn threatens project sustainability and the health of the open source ecosystem.

Empirically studying toxicity on GitHub is also methodologically challenging. Toxicity in conversations is shaped by interaction structure and context, not surface text alone~\cite{anuchitanukul2022revisiting}. Moreover, toxicity detection remains inherently subjective: judgments vary across annotators, perspectives, and context, including when LLMs are used as evaluators~\cite{atil2026robust}. As a result, even frontier LLMs do not reliably detect GitHub toxicity. For instance, when we evaluated Nemotron-3-Super~\cite{blakeman2025nvidia}, a large frontier model with 120B parameters, on Imran et al.'s~\cite{imran2026toxicity} dataset of 898 human-annotated GitHub conversations, it correctly classified all 696 non-toxic instances but missed half of the toxic ones (detecting only 101 out of 202), revealing a strong bias toward predicting non-toxicity. Such results suggest that generic model judgments miss the implicit, context-dependent forms of toxicity common in software engineering discourse.

Several well-known studies have examined toxicity on GitHub, but most have been small in scale. Miller et al.~\cite{miller_did_2022} analyzed 100 issues and found that toxicity is often directed at individuals rather than code, and appears more frequently in discussions involving repeated reporters. Ferreira et al.~\cite{ferreira_how_2022} annotated 205 locked issues and observed that many toxic discussions emerge from miscommunication or disagreement about code and tend to escalate over time. These studies identified important patterns, but their limited scale constrains generalization.

Scaling this line of work is difficult because strong indicators of toxicity are rare. For example, locked issues account for only 0.22\% of GitHub issues~\cite{raman2020stress}. This creates a two-sided gap: the field lacks large-scale, GitHub-calibrated toxicity labels, and producing such labels is expensive. Full manual annotation is prohibitively costly, while generic LLM labeling can be unreliable when toxicity depends on community-specific norms and discussion context~\cite{kolla_llm_mod_2024}.

Recent work shows that LLMs can support toxicity detection on GitHub~\cite{mishra_exploring_2024, anindya2026toxishield} and can help model conversational dynamics for forecasting derailment in OSS discussions~\cite{imran2026toxicity}. At the same time, these studies show that performance depends heavily on prompt design, contextual framing, and task formulation, rather than on simply applying a powerful model out of the box. In addition, frontier LLMs can be expensive to deploy at scale, making annotation efficiency a methodological requirement, not just an engineering concern.

In this paper, we address this gap with two contributions: a methodological one and an empirical one. On the methodological side, we introduce a human-in-the-loop (HITL) annotation pipeline that produces GitHub-calibrated toxicity labels at low cost. A single call to a small, local LLM assigns a first-pass toxicity label and simultaneously produces a set of interpretable event category scores that characterize the conversation’s tone. A lightweight Random Forest validator then uses those scores to flag the small subset of conversations most likely to be mislabeled, sending only those for human review. This design avoids repeated prompting and multi-pass labeling, reduces annotation cost, and adapts to GitHub-specific communication patterns, building on HITL methodology from Wang et al.~\cite{wang2024human}.

On the empirical side, we apply this pipeline to annotate over 124K GitHub issue and pull request discussions for toxicity. The paper is organized accordingly: Section 2 describes the pipeline’s design, implementation, and evaluation, while Sections 3--5 use the resulting dataset to revisit prior findings at scale and present new insights into toxicity dynamics.
The main contributions of this paper are:

\begin{itemize}[leftmargin=*,topsep=0pt]
\item Development and evaluation of a HITL annotation pipeline for GitHub toxicity that combines single-pass local-model annotation with a lightweight validation model, requiring only one LLM call per conversation and reducing human review to roughly 1.5\% of the dataset.
\item A large-scale dataset of over 124K GitHub issue and pull request discussions annotated for toxicity.
\item Empirical validation of toxicity hypotheses from prior small-scale studies on a substantially larger dataset.
\item New insights into toxicity dynamics, including project governance, contributor behavior, and language ecosystem differences.
\end{itemize}

Our findings can inform the design of automated moderation tools, improve community guidelines, and support initiatives that promote a healthier open source environment. The dataset and methodology can also serve as a foundation for future research in computational social science and software engineering.

\section{Human-in-the-Loop Annotation of Toxicity on GitHub}

Annotating GitHub discussions for toxicity at scale requires balancing automation with reliability. LLMs can provide efficient first-pass labels, but their outputs may be inaccurate or systematically biased relative to human judgment~\cite{zhang-etal-2024-sentiment,liu-etal-2024-emollms,imran_2024_emocause,zhang2020effect,zhu2024exploring}, motivating hybrid strategies that combine automated labeling with targeted human review~\cite{wang2024human,pangakis2025keeping}.

This section presents our human-in-the-loop annotation methodology. To avoid circularity, we develop and evaluate the validator on previously annotated GitHub datasets before applying it to our newly collected large-scale 2024 corpus (Section~3).

Our pipeline has three stages. First, an LLM generates an initial binary toxicity prediction along with event category scores that characterize the conversation's tone and dynamics (\S\ref{sec:first-pass}). A separate machine learning validator then estimates whether the LLM output is likely incorrect, using the event category scores to identify unreliable annotations (\S\ref{sec:annot_model}). Finally, human annotators review only those instances flagged by the validator, concentrating manual effort on the conversations most likely to be mislabeled.

\added{Our pipeline is thus human-in-the-loop in the annotation-verification sense established in prior work on human–LLM collaborative labeling~\cite{pangakis2025keeping,wang2024human}. The LLM is not retrained during annotation, rather, human judgment enters the pipeline at two points: 1) human-annotated data trains the validator that decides which LLM outputs require review, and 2) human annotators serve as the final label authority for every flagged instance.}

\subsection{First-Pass LLM Toxicity Annotation}\label{sec:first-pass}

The first stage of the pipeline assigns an initial toxicity label to each GitHub conversation using the Ministral model ({\em ministral3:14b})~\cite{liu2026ministral}. We selected this model because it is open weight under Apache 2.0, supports long contexts of up to 128k tokens, and is small enough to process large collections of conversations efficiently on available hardware. Larger alternatives such as \textit{Llama 3.3:70B} were impractically slow in our setting, whereas smaller alternatives such as \textit{Ministral-3:8B} were less accurate.

With temperature set to 0, the model produces two primary outputs for each conversation: a binary toxicity prediction (1 if clear toxic language is present, 0 otherwise) and a set of eight event category scores (described below) that decompose the conversation along dimensions known to make toxicity annotation difficult in software engineering contexts. The event category scores are not used to classify toxicity directly; rather, they serve as features for a separate validation model that identifies conversations where the LLM is likely to produce unreliable labels (Section~\ref{sec:baseline-comparison}). We also use another prompt to detect the toxicity prediction with a confidence score (0--10) for the generated toxicity and a brief explanation, retained for baseline comparisons.

\subsubsection{Prompt Structure}

The prompt guides the model through a step-by-step toxicity assessment with three main elements:
\begin{itemize}[leftmargin=*]
\item \textbf{Toxicity Definition:} A definition tailored to GitHub conversations, drawing on the toxicity categories of Miller et al.~\cite{miller_did_2022} (e.g., insulting, entitled, arrogant) and the incivility markers of Ferreira et al.~\cite{ferreira2021shut} and Ehsani et al.~\cite{ehsani2024incivility} (e.g., frustration, mocking, dismissiveness). A conversation is toxic if it contains rude, disrespectful, dismissive, antagonistic, profane, mocking, or hostile language that could make a reasonable contributor feel unwelcome. The definition distinguishes language directed at people from strong negative language directed at project artifacts~\cite{miller_did_2022,sarker_automated_2023}, noting that harsh technical criticism can be non-toxic if it remains professional.
\item \textbf{Binary Classification Task:} Instructions to assign a toxicity prediction (0/1), a confidence score (0--10), and a brief explanation.
\item \textbf{Event Category Scoring:} Instructions to score each of eight event categories from 0 to 10 based on explicit text evidence. These categories (Table~\ref{tab:event-categories}) were derived from prior studies of toxicity, incivility, and conflict in open source communities, and capture dimensions along which conversations become difficult to annotate---e.g., strong process criticism mixed with resolution language, or escalating tension without overt abuse. The prompt emphasizes that category scoring and toxicity prediction are separate tasks, and that a high score on any single category does not automatically indicate toxicity.
\end{itemize}

% The following example illustrates the model’s output for a toxic GitHub conversation.

% \promptbox{Example Toxic Conversation}{
% {\bf commenter1:} If you set up some task this way: before deploy’, some\_task’, And in some\_task you simply read the “latest\_release” variable (in a puts line for instance), then:
% Current results: [...]\\
% This is a big f***up IMHO.\\
% {\bf commenter2:} [\ldots] Profanity aside, you may have a point. I’ll take a patch.\\
% {\bf commenter1:} Why do you close the issue?\\
% {\bf commenter2:} Because you don’t offer a patch, and profanity really pisses me off.\\
% {\bf commenter1:} Mutable programming pisses me off [\ldots]
% }

% \promptbox{LLM Output}{
% {\bf toxicity\_prediction:} 1,\\
% {\bf confidence:} 7,\\
% {\bf explanation\_text:} ``The conversation contains profanity and hostile exchanges. The phrase `big f***up’ and responses such as `profanity really pisses me off’ create an unwelcoming tone. The closing of the issue without explanation is dismissive.’’\\
% {\bf event\_category\_scores:} person\_hostility: 3, process\_hostility: 6, dismissiveness: 5, escalation: 4, subtle\_criticism: 2, hostility\_density: 3, resolution\_language: 1, behavior\_callout: 2
% }

\subsection{Annotation Validation Model}\label{sec:annot_model}

Although the first-pass LLM outputs were generally reasonable, they remained prone to error in cases common in software engineering discourse, such as project-specific figurative language, lengthy discussions containing stack traces or logs, and emotionally charged technical exchanges that are not necessarily toxic~\cite{imran_2024_emocause}. We therefore introduce a separate validation model that identifies conversations where the LLM annotation is most likely to be unreliable. Crucially, the validator does not relabel conversations or detect toxicity directly. Instead, it learns to recognize \textit{confusing} conversations---those where competing signals (e.g., strong process criticism alongside resolution language, or subtle hostility mixed with collaborative tone) make the correct label hard to determine automatically---and flags them for human review.

\subsubsection{Validator Features}

The validator uses the eight event category scores produced by the LLM prompt as its feature set. These scores, each ranging from 0 to 10, characterize different dimensions of the conversation’s tone and interpersonal dynamics. Table~\ref{tab:event-categories} summarizes the categories and the prior work motivating each one.

\begin{table}[t]
    \small
    \caption{Event category features scored by the LLM (0--10), used by the validator to identify conversations likely to be hard for the LLM to annotate.}
    \centering
    \setlength{\tabcolsep}{3pt}
    \begin{tabular}{p{1.5cm} p{6.5cm}}
        \hline
        \textbf{Category} & \textbf{Definition} \\
        \hline
        \raggedright\texttt{person\_\allowbreak hostility}
        & Hostile language directed at a person or group (insults, name-calling), excluding criticism of code or process~\cite{miller_did_2022,sarker_automated_2023,gunawardena2022destructive}. \\ \hline
        % \addlinespace[4pt]
        \raggedright\texttt{process\_\allowbreak hostility}
        & Strongly negative language directed at code, tools, processes, or decisions rather than people. Common in OSS and largely independent of toxicity~\cite{miller_did_2022,sarker_automated_2023}. \\ \hline
        % \addlinespace[4pt]
        \raggedright\texttt{dismiss\-iveness}
        & Language that shuts down or refuses to engage with another participant’s contribution or concern~\cite{ferreira2021shut,trinkenreich2023belong}. \\ \hline
        % \addlinespace[4pt]
        \raggedright\texttt{escalation}
        & Conversation becomes more tense or confrontational over time without overt insults, slurs, or profanity~\cite{miller_did_2022,ehsani2024incivility,ferreira_how_2022}. \\ \hline
        % \addlinespace[4pt]
        \raggedright\texttt{subtle\_\allowbreak criticism}
        & Mild or indirect criticism or disapproval that is noticeable but not strongly explicit~\cite{ehsani2024incivility,ferreira_how_2022}. \\ \hline
        % \addlinespace[4pt]
        \raggedright\texttt{hostility\_\allowbreak density}
        & Concentration of person-directed hostile language relative to conversation length~\cite{saveski_structure_2021}. \\ \hline
        % \addlinespace[4pt]
        \raggedright\texttt{resolution\_\allowbreak language}
        & Signs of agreement, de-escalation, or collaborative closure, indicating heated exchanges were normal debate~\cite{filippova2016effects}. \\ \hline
        % \addlinespace[4pt]
        \raggedright\texttt{behavior\_\allowbreak callout}
        & Commentary explicitly naming problematic behavior (e.g., ``that’s rude’’), indicating community self-policing~\cite{ferreira2024incivility}. \\
        \hline
    \end{tabular}
    \label{tab:event-categories}
\end{table}

Using the event category scores as validator features offers several advantages over externally computed features. The scores are derived directly from the LLM’s reading of the conversation, so they reflect the same evidence the model used for its toxicity prediction. They decompose the conversation into interpretable dimensions that can reveal \textit{conflicting signals}, for example, a conversation with high process hostility but also high resolution language is inherently ambiguous and more likely to be misclassified. Because the scores are produced alongside the toxicity prediction in a single LLM call, they add no additional inference cost.

\subsubsection{Training Procedure}
We trained a Random Forest validation model on Imran et al.'s manually annotated 898 GitHub conversations (202 toxic, 696 non-toxic)~\cite{imran2026toxicity}.  Because the dataset is imbalanced (approximately 22\% toxic), we applied SMOTE (Synthetic Minority Over-sampling Technique) to the training folds to balance the class distribution during model training. We evaluated the model using stratified 5-fold cross-validation to preserve the class distribution in each fold and to obtain more robust performance estimates across the full dataset. We implemented the model using the {\em scikit-learn} and {\em imbalanced-learn} libraries~\cite{scikit-learn}.

\subsubsection{Validator Evaluation} Table~\ref{tab:validator-perf} presents the validator’s 5-fold cross-validation performance at different thresholds~$\tau$, reporting metrics for the \textit{Incorrect} class (LLM~$\neq$~Human) and \textit{Correct} class (LLM~$=$~Human). High Incorrect recall is desirable, as it indicates the validator effectively flags LLM mistakes for human review. As $\tau$ increases, Incorrect Recall rises while Precision decreases, reflecting a trade-off between catching errors and minimizing unnecessary review. At $\tau = 0.5$, the model achieves the best balance: Macro F1 of $0.83$, Incorrect F1 of $0.68$, and Weighted F1 of $0.94$.

To understand which features are most informative for the validator, we computed Random Forest feature importances (mean decrease in impurity) averaged across the cross-validation folds. The most influential feature is \texttt{escalation} (importance 0.373), followed by \texttt{subtle\_criticism} (0.249) and \texttt{process\_hostility} (0.158). Together, these three features account for over 78\% of the total importance, indicating that the validator relies most heavily on signals of indirect tension and tone rather than overt hostility. Notably, \texttt{person\_hostility} (0.039) and \texttt{behavior\_callout} (0.006) contribute the least, suggesting that conversations with clear interpersonal hostility or explicit norm enforcement are relatively easy for the LLM to classify correctly, whereas ambiguity arises primarily in conversations with escalating but implicit tension.

\begin{table}[t]
    \small
    \caption{Validator performance (5-fold cross-validation) at selected thresholds~($\tau$). \textit{Inc.}~=~Incorrect class (LLM~$\neq$~Human); \textit{Cor.}~=~Correct class.}
    \centering
    \setlength{\tabcolsep}{3.5pt}
    \begin{tabular}{r | ccc | ccc | cc}
        \hline
        & \multicolumn{3}{c}{\bf Inc.} & \multicolumn{3}{c}{\bf Cor.} & {\bf Macro} & {\bf Weighted} \\
        \cline{2-4} \cline{5-7}
        {\bf $\tau$}
        & {\bf P} & {\bf R} & {\bf F1}
        & {\bf P} & {\bf R} & {\bf F1}
        & {\bf F1} & {\bf F1} \\
        \hline
        0.1 & 0.76 & 0.32 & 0.45 & 0.94 & 0.99 & 0.96 & 0.71 & 0.92 \\
        0.3 & 0.70 & 0.54 & 0.61 & 0.96 & 0.98 & 0.97 & 0.79 & 0.94 \\
        \rowcolor{gray!20} 0.5 & 0.68 & 0.69 & 0.68 & 0.97 & 0.97 & 0.97 & 0.83 & 0.94 \\
        0.7 & 0.60 & 0.76 & 0.67 & 0.98 & 0.95 & 0.96 & 0.82 & 0.94 \\
        0.9 & 0.42 & 0.86 & 0.56 & 0.99 & 0.89 & 0.93 & 0.75 & 0.90 \\
        \hline
    \end{tabular}
    \label{tab:validator-perf}
\end{table}

\subsubsection{Baseline Comparison}\label{sec:baseline-comparison}

To assess whether the event category features genuinely help identify difficult-to-classify conversations, we compare our validator against several baselines (Table~\ref{tab:baseline-comparison}). The three Random Forest variants share the same setup: SMOTE oversampling and stratified 5-fold cross-validation on the 898-conversation dataset, with the best threshold reported. The remaining baselines are evaluated on the full dataset without cross-validation, since they involve no trained model.

\paragraph{Confidence Threshold.} The simplest baseline flags predictions whose self-reported confidence falls below a threshold. This performed poorly: for thresholds 0-7 it flagged no incorrect predictions (Macro F1~$=$~$0.488$, equivalent to the majority class). Even at the best threshold of 9, Macro F1 reached only $0.592$ with Incorrect Recall of $0.143$, flagging just 15 of 898 instances and missing most LLM errors---indicating that the model's confidence is poorly calibrated for this task.

\paragraph{RF on Confidence + Prediction.} To test whether a learned decision boundary helps, we trained a Random Forest on the confidence score and binary prediction. This achieved a Macro F1 of $0.64$ at $\tau=0.5$, flagging 101 of 898 instances---an improvement over simple thresholding, but with weak Incorrect-class performance (F1~$=$~$0.34$). The model tends to overfit on the binary prediction: because the prediction is correct most of the time, the Random Forest relies on it as a shortcut rather than learning the conversational patterns that characterize genuinely difficult cases.

\paragraph{RF on All 10 Features.} Adding the eight event category scores to the confidence and prediction (10 features total) tests whether these direct LLM outputs carry complementary information. This model reaches Macro F1 of $0.80$ at $\tau=0.5$, flagging 88 instances---but still underperforms our validator (Macro F1~$=$~$0.83$, 80 instances). The same overfitting dynamic persists: the Random Forest assigns disproportionate importance to the prediction feature, which acts as a proxy for the majority class and crowds out the subtler category-based signals most informative for detecting annotation difficulty.

\paragraph{Second LLM Disagreement.} We also tested flagging conversations where two LLMs disagree. We ran Qwen ({\em Qwen3-14b})~\cite{yang2025qwen3}, an instruction-tuned model with a different architecture, on the same 898 conversations using the same prompt. Despite doubling the inference cost, this baseline performed worst overall (Macro F1~$=$~$0.476$, Incorrect F1~$=$~$0.048$): although Qwen disagreed with Ministral on 87 conversations, this disagreement failed to identify most incorrect annotations.

\begin{table*}[t]
    \small
    \caption{Comparison of validator approaches for identifying incorrect LLM annotations. RF models use SMOTE and stratified 5-fold cross-validation; other baselines are evaluated on the full 898-conversation dataset. Best threshold per approach is shown. \textit{LLM Calls}: inference calls per conversation. \textit{Human Review}: estimated percentage of the dataset requiring manual annotation.}
    \centering
    \begin{tabular}{l|c|c c c|c c c|c|c|c}
        \hline
        \multirow{2}{*}{\bf Approach}
        & \multirow{2}{*}{\bf $\tau$}
        & \multicolumn{3}{c|}{\bf Incorrect (LLM $\neq$ Human)}
        & \multicolumn{3}{c|}{\bf Correct (LLM $=$ Human)}
        & {\bf Macro}
        & {\bf LLM}
        & {\bf Human} \\
        & & {\bf Prec.} & {\bf Rec.} & {\bf F1}
        & {\bf Prec.} & {\bf Rec.} & {\bf F1}
        & {\bf F1}
        & {\bf Calls}
        & {\bf Review (\%)} \\
        \hline
        Confidence threshold & 9 & 0.400 & 0.143 & 0.211 & 0.959 & 0.989 & 0.974 & 0.592 & 1 & 1.7\% (15/898) \\
        RF (conf.\ + pred.) & 0.5 & 0.238 & 0.571 & 0.336 & 0.977 & 0.910 & 0.943 & 0.639 & 1 & 11.2\% (101/898) \\
        RF (all 10 features) & 0.5 & 0.602 & 0.679 & 0.639 & 0.969 & 0.957 & 0.963 & 0.801 & 1 & 9.8\% (88/898) \\
        2nd LLM disagreement & n/a & 0.046 & 0.051 & 0.048 & 0.909 & 0.899 & 0.904 & 0.476 & 2 & 9.7\% (87/898) \\
        \hline
        \bf Validator (ours) & \bf 0.5 & \bf 0.675 & \bf 0.692 & \bf 0.684 & \bf 0.971 & \bf 0.968 & \bf 0.969 & \bf 0.827 & \bf 1 & \bf 8.9\% (80/898) \\
        \hline
    \end{tabular}
    \label{tab:baseline-comparison}
\end{table*}

Table~\ref{tab:baseline-comparison} confirms that raw confidence is a weak signal: the LLM tends to express high confidence even when wrong, likely because toxic and non-toxic software engineering conversations share surface features (e.g., strong language about code). Including the prediction and confidence alongside the category scores does not help---these features dilute the category-based signal because the classifier overfits on the prediction, which is correct roughly 90\% of the time and dominates the learned splits. Using the eight category scores alone yields the strongest results (Macro F1~$=$~$0.827$, Incorrect Recall~$=$~$0.692$), flagging nearly five times as many LLM errors as the confidence threshold while requiring no additional inference cost.

\added{\subsubsection{Validator Independence.} The performance drop when prediction and confidence are included with the event scores (Macro F1 = 0.801) compared to using the event scores alone (Macro F1 = 0.827) suggests that the event scores capture information about annotation difficulty beyond the LLM’s direct toxicity judgment. Although both outputs come from the same model, they answer structurally different questions. The prediction provides a direct binary judgment of toxicity, whereas the event scores decompose the conversation into dimensions that can make that judgment difficult, such as strong process criticism paired with resolution language. The feature importance analysis in Section 2.2.3 supports this interpretation: the validator relies primarily on indirect signals of tension, especially escalation and subtle criticism, rather than on overt hostility features most aligned with the toxicity label. This indicates that the validator succeeds by learning which conversational patterns make labeling hard, not simply by reproducing the original prediction.}

\subsubsection{Pipeline Annotation Cost}\label{sec:cost}

A key advantage of our pipeline is its low annotation cost. Each conversation requires exactly one LLM call that produces the toxicity prediction and all eight event category scores in a single structured output. The Random Forest validator adds negligible overhead---training takes under one second and scoring the full \totalCount-conversation corpus takes only seconds on commodity hardware. Human effort is concentrated on the small fraction of conversations the validator flags; on our large-scale corpus, this amounted to roughly 1.5\% of all conversations (Section~3). As the \textit{Human Review} column in Table~\ref{tab:baseline-comparison} shows, our validator achieves the best Macro F1 while keeping human annotation effort minimal---the same single-call inference budget as the confidence threshold, but with substantially better error detection.

% \subsubsection{Validator Error Analysis}\label{sec:error-analysis}
% Having established the validator's quantitative performance, we now examine its failure modes qualitatively. Two of the authors manually examined instances where the validator failed to identify LLM mistakes across the cross-validation folds. We analyzed the event category scores alongside Ministral’s reasoning for these cases.

% In 5 of the 6 cases, the ground truth annotations were highly subjective. In some, we agreed with Ministral’s assessment over that of the human annotators, consistent with prior research showing that annotations involving emotion, sentiment, and politeness can vary significantly across evaluators~\cite{BOCK2025112339, imtiaz2018}. The remaining case involved capitalization in comments, which the model interpreted as a toxicity signal but the annotators did not; we agreed with the annotators but found the model’s reasoning understandable.

% Since most errors stem from subjective judgments or mild toxicity that was later resolved, we conclude that the validator’s performance and the resulting dataset remain reliable for practical use.

% %To further assess this, we examined the 13 True Incorrect cases and found that the validator consistently identified clear mistakes in the LLM’s predictions.

\subsection{Generalization to External Dataset}

To evaluate generalization, we applied our pipeline to the dataset of Raman et al.~\cite{raman2020stress}, which has no overlap with Imran et al.’s data (different sampling strategy and time period, 2012-2018). Raman et al. targeted GitHub issues locked as ``too heated’’ or containing maintainer reactions referencing ``attitude.’’ We used the 314 instances (71 toxic, 243 non-toxic) available at the comment level~\cite{toxicity_detector_dataset}.

On these 314 conversations, Ministral predicted 297 correctly, yielding a weighted F1 of 0.96. Of the 17 incorrect predictions, the validator flagged 10 for review. The remaining 7 undetected errors---3 of which involved toxic conversations---all exhibited
% the same pattern found in our earlier error analysis (Section~\ref{sec:error-analysis}):
subjective or borderline toxicity where annotator judgment could reasonably differ (e.g.,~\cite{forgottenserver_issue2494}, annotated as non-toxic, which is debatable). These findings suggest that the HITL methodology generalizes effectively across datasets with differing sampling and annotation criteria.

\section{Large-Scale Data Collection and Annotation}

Having established and evaluated the annotation pipeline, we now apply it to a newly collected large-scale GitHub corpus. This section describes the construction of the 2024 dataset used in the remainder of the paper: repository sampling and conversation collection, application of the first-pass LLM annotator and validator, and production of the final labeled dataset.

% \begin{figure*}[t]
% \centering
% \includegraphics[width=0.99\linewidth]{large-scale-toxicity.pdf}
% \caption{Overview of the HITL annotation pipeline applied to the 2024 GitHub corpus. Stage~1 (LLM annotation) and Stage~2 (validation model) are described in Section~2; Stage~3 (targeted manual review) and the resulting dataset are described in this section.}
% \label{fig:overview}
% \end{figure*}

\subsection{Large-Scale Corpus Collection}

To construct a broad and realistic sample of GitHub discussions, we collected conversations from active, collaborative, and software-relevant repositories. Our sampling strategy follows the guidance of Kalliamvakou et al., who argue that empirical studies on GitHub should prioritize repositories with meaningful collaborative activity while avoiding personal, inactive, or non-software projects~\cite{promisesperils2014}.

We began by selecting repositories with at least 1,000 stars as a signal of community interest, at least 100 total issues (open or closed), and at least one commit within the 10 months preceding our collection window starting 1 January 2024. We further required that each repository have either a locked conversation or an explicit code of conduct, since these suggest active moderation: locked conversations indicate maintainer intervention in heated discussions, while a code of conduct reflects an explicit attempt to define acceptable behavior. The first author then manually reviewed the candidate list and excluded clearly non-software repositories, such as programming problem collections and curated lists.

Using the GitHub GraphQL API~\cite{github_graphql}, we collected all public English-language issue and pull request conversations created from January 2024 through January 2025 from each of the 340 selected repositories. We retained only conversations with at least one comment beyond the initial description (i.e., at least two total comments), ensuring the corpus captures actual interaction rather than single-post reports.

Table~\ref{tab:github-stats} summarizes the resulting corpus. We collected \totalCount conversations from 340 repositories, comprising 39,891 pull request and 84,866 issue discussions. Conversation length varies substantially, from very short exchanges to threads of up to 29,078 words, with a median of 3 comments and 41 words.

\begin{table}[tb]
\small
\caption{Statistics of the GitHub conversations sample.}
\centering
\begin{tabular}{l | l}
\hline
\textbf{Metric} & \textbf{Value} \\
\hline
Total Repositories & 340 \\
\hline
Pull Request Conversations & 39,891 \\
Issue Conversations & 84,866 \\
Total Sampled Conversations & \totalCount \\
% \midrule
% Total Annotated Conversations & \totalCount \\
\hline
Median Comments per Conv. & 3 \\
Median Conversation Length & 41 words \\
% Minimum Conversation Length & 1 word \\
Maximum Conversation Length & 29,078 words \\
\hline
Time Frame Start & 2024-01-01 00:54:06 UTC \\
Time Frame End & 2025-01-31 23:47:57 UTC \\
Total Days & 396 \\
\hline
\end{tabular}
\label{tab:github-stats}
\end{table}

\added{The \totalCount{} conversations involve 100,907 unique participants.
For 10,247 of them, author association data was not available.
Of the remaining 90,660, 88,674 carry exactly one role. Among them, 76,466 appear as \textsc{None} (external, unaffiliated participants), 10,550 as Contributors, 1,338 as Members, 289 as Collaborators, and 31 as Owners.
The remaining 1,986 carry multiple roles, reflecting that GitHub author associations are repository-specific and a participant's role may differ across repositories. For example, an Owner of one repository may appear as \textsc{None} when commenting in another. 
% The most common combinations are Contributor+None (1,560), Member+None (171), and Contributor+Member (102).
}

\subsection{Applying the HITL Annotation Pipeline}

We applied the pipeline described in Section~2 to the collected corpus. Because the validator was trained on Imran et al.'s dataset, which was sampled from GitHub prior to our 2024 collection window, there is no overlap between the training data and the corpus annotated here. Each conversation was processed by the Ministral-based annotator, producing a binary toxicity prediction along with event category scores.

% Before annotation, we pre-processed conversation transcripts by removing code snippets, output and error logs, stack traces, and URLs. These elements are common in GitHub discussions but can distract the model from the interpersonal content relevant to toxicity assessment. We also found that LLM-specific terms such as ``prompt’’ and ``prompting’’ occasionally confused the annotator, so we replaced them with a neutral token during preprocessing.

% The first-pass annotation completed successfully for nearly the entire corpus; we excluded one conversation for which Ministral returned malformed JSON. After this exclusion, the annotated corpus contained \totalCount conversations.

We then ran the Random Forest validator on the LLM-labeled corpus using the threshold selected in Section~\ref{sec:annot_model} ($\tau = 0.5$). The validator flagged 1,874 conversations (approximately 1.5\% of the corpus) for manual review.

% \subsection{Targeted Manual Review}

Two of the authors, \removed{both with prior experience studying toxicity in GitHub discussions}\added{one a Ph.D.\ holder in software engineering and one a Ph.D.\ candidate, both with prior experience studying toxicity in GitHub conversations}, manually reviewed the 1,874 flagged conversations. \removed{For each conversation, the annotator examined the discussion alongside the Ministral-generated toxicity prediction and explanation, then determined whether the initial label should be retained.}\added{They used toxicity definitions, category taxonomies, and examples from prior GitHub toxicity studies (Miller et al.~\cite{miller_did_2022}, Ferreira et al.~\cite{ferreira2021shut}, Imran et al.~\cite{imran2026toxicity}, and Raman et al.~\cite{raman2020stress}) to calibrate their judgments. For each flagged conversation, the assigned annotator reviewed the discussion together with the Ministral prediction and explanation, assigned a binary toxicity label, and recorded a brief justification.} The manual review required approximately 16 person hours of effort. %Without the validator, reviewing the entire corpus of \totalCount conversations would have required an estimated ~1650 person-hours---a reduction of approximately 99.03\% in human annotation effort.

To assess consistency between the two annotators, we conducted an inter-annotator agreement study on a shared subset of 200 flagged conversations, drawn from the 1,874, that both annotators reviewed independently. The resulting Cohen's $\kappa$ was 0.78, indicating substantial agreement. Disagreements were resolved through discussion, justifying the division of the remaining 1,674 flagged conversations equally between the two annotators for individual review.

Manual review led to 169 label changes. Ministral had originally classified 805/124,757 conversations as toxic and 123,952/124,757 as non-toxic. The corrections were predominantly in one direction: 155 conversations were relabeled from non-toxic to toxic, while 14 were changed from toxic to non-toxic, indicating that the LLM's primary failure mode among flagged conversations was under-detection of toxicity. The two annotators made 78 and 91 corrections respectively, suggesting that the flagged cases were distributed relatively evenly across the two subsets.

\subsection{Validation on Unflagged Conversations}

The validator is designed to flag conversations where the LLM is likely to have made an error, but it may also miss some errors among the unflagged majority. To estimate the rate of such missed errors, we drew a stratified random sample of 5,000 conversations, comprising 4,800 conversations predicted non-toxic by the model and 200 predicted toxic, from the 122,883 conversations that the validator did \textit{not} flag for review. %We stratified the sample by the LLM's toxicity prediction to ensure adequate representation of both predicted-toxic and predicted-non-toxic conversations.

Two annotators independently reviewed an initial subset of 500 sampled conversations using the same annotation protocol applied to the flagged set. They achieved a $\kappa$ of 0.978 on this subset, then resolved disagreements through discussion and reached full agreement. The remaining conversations were then divided between the annotators for individual review. For each conversation, the annotator determined whether the model's original toxicity label was correct.

Across the 5,000 unflagged conversations, the reviewers found 11 labeling errors in total. Among the 4,800 conversations predicted non-toxic by the LLM, 4 were actually toxic, yielding an LLM false-negative rate of 0.083\% with a 95\% confidence interval of 0.023\% to 0.213\%. Among the 200 conversations predicted toxic by the LLM, 7 were actually non-toxic, yielding an LLM false-positive rate of 3.5\% in the predicted-toxic stratum, with a 95\% confidence interval of 1.42\% to 7.07\%. Overall, this suggests that the validator misses very few toxic conversations among the unflagged majority. While the point estimate of the false-positive rate is 3.5\%, the wide confidence interval (1.42\% to 7.07\%) indicates that this estimate should be interpreted with caution, as it is based on a small number of observed misclassifications (n = 7). Given the extreme rarity of toxic content in GitHub conversations and the resulting class imbalance, these results suggest that the validator performs reasonably well on the unflagged set. 

In 3 of the 4 false-negative cases (toxic threads mislabeled as non-toxic), the conversation involved a disagreement in tone around an issue or PR that was closed without resolution. The 7 false-positive cases (non-toxic threads mislabeled as toxic) were more varied: three had comments where suggestions were discarded, two involved a user expressing impatience over an unresolved bug, one contained spam comments, and one involved name-calling.

% \todo{Describe the nature of the errors---e.g., ``All were borderline cases involving mild dismissiveness or ambiguous tone, consistent with the patterns observed in the validator's error analysis (Section~2).''}

Extrapolating the observed error rates from the manual recheck to the full set of unflagged conversations yields an estimated 129 additional labeling errors in the corpus beyond those corrected through the validator, for an estimated overall labeling accuracy of 99.90\%.
These results indicate that the validator successfully concentrates the majority of LLM errors into the flagged set, with a residual error rate of 0.105\% among unflagged conversations.  %Given the extreme class imbalance and low base rate of toxicity in GitHub discussions, this level of residual error is sufficiently low to support reliable downstream analysis.

% \subsection{Comparison with a State-of-the-Art Detector}
\added{
As an additional baseline, we evaluated ToxiShield, a recent SE-specific toxicity detector whose BERT-based module outperformed prior techniques~\cite{anindya2026toxishield}. We trained the detection module with the authors' public code and dataset, using BERT as the backbone and following their original setup. We then applied the trained model to our manually verified sample of 5,000 unflagged GitHub conversations. Because our complete pipeline includes targeted human correction, it is not directly comparable to fully automated toxicity detectors. We therefore compare ToxiShield with the automated portion of our pipeline, before human correction, as a positioning baseline rather than a full detector benchmark.
}

\added{
ToxiShield achieved a toxic-class F1 of 0.148, with macro, micro, and weighted F1 scores of 0.531, 0.843, and 0.803, respectively. In contrast, the automated portion of our HITL pipeline, consisting of Ministral annotation followed by validator-based filtering with no human correction, achieved a toxic-class F1 of 0.972 and macro, micro, and weighted F1 scores of 0.986, 0.998, and 0.998. ToxiShield tended to overpredict toxicity, often flagging isolated frustration or negative wording without considering the surrounding discussion context. This pattern is consistent with its reported error modes, including context misreads, self-deprecating humor, and technical jargon or inline code interpreted as hostile~\cite{anindya2026toxishield}.
% For example, ToxiShield flagged ``\textit{now i have to figure out how to get the blasted macos app to not kill itself upon launching}'' and ``\textit{We have tried every sensible and stupid solution, playing around with rendering and parameter set events but have not been able to make this happen no matter what we do.}'' Our approach did not label these conversations as toxic. We interpret this gap cautiously as ToxiShield was trained on code review comments and classifies individual comments, whereas our task labels complete GitHub conversations.
}

% \added{
% }

\subsection{Final Dataset}

After manual correction, the final dataset contains \totalCount conversations: \totaltoxic labeled toxic and 123,811 labeled non-toxic. This corpus serves as the basis for the empirical analyses in the remainder of the paper.

The large-scale deployment confirms the practical value of the HITL pipeline: the first-pass annotator provided efficient coverage of the full corpus, the validator reduced the review burden to roughly 1.5\% of conversations, and targeted manual review corrected the residual errors.

\section{Validating Prior Findings at Scale}

In this section, we use our large-scale toxicity dataset to test hypotheses introduced in prior work on toxicity in software engineering and online discussions. We selected these papers because they present empirically grounded, testable claims about toxicity dynamics, many of which were based on smaller datasets or limited contexts. Raman et al.~\cite{raman2020stress} and Miller et al.~\cite{miller_did_2022} analyzed toxicity trends on GitHub. Imran et al.~\cite{imran2026toxicity} and Sarker et al.~\cite{sarker2025landscape} introduced annotated datasets of toxic GitHub discussions. Lastly, Saveski et al.~\cite{saveski_structure_2021} provided structural insights from toxicity on X (Twitter), which we treat as transferable to GitHub’s threaded discussions. 

\subsection{Participants of Toxic Conversations}\label{sub:participants}

Prior work by Miller et al.~\cite{miller_did_2022} and Imran et al.~\cite{imran2026toxicity} found that external participants (e.g., users who are not developers of the project), rather than active project contributors, play a significant role in driving toxicity in GitHub issue/PR discussions. Imran et al., using a dataset of 202 toxic and 696 non-toxic threads, found that 76.0\% of toxic conversations were initiated by external participants and made 52.79\% of the comments in toxic threads. In contrast, non-toxic discussions had a higher engagement from project contributors (66.62\%), indicating their role in fostering constructive discussions. 

GitHub assigns roles to participants in issue/PR discussions based on their involvement level within a repository. We used the following five categories of GitHub participants that we found in our corpus: None, Member, Contributor, Collaborator, and Owner~\cite{github_comment_author_association}. 
Participants labeled as "None" have no affiliation with the project, while Members belong to the owning organization. Contributors are people who have previously committed code, Collaborators are invited contributors, and Owners have administrative control. Following Imran et al.'s approach, we grouped \textit{Contributor, Collaborator, Member,} and \textit{Owner} as project contributors, and \textit{None} participants as external participants, i.e., unaffiliated with the repository.

We analyzed our dataset of {\totalCount} GitHub issue/PR conversations and categorized participants by their level of involvement in a repository. Overall, external participants contributed 44.48\% of all comments, while project contributors accounted for 55.52\%. In toxic conversations, however, this distribution shifts: external participants contributed 57.56\% of comments, while project contributors contributed only 42.44\%. When examining toxicity levels by the proportion of project contributor comments in a discussion, we found a non-monotonic pattern (Table~\ref{tab:developer_toxicity}). Toxicity initially rises as contributor participation increases, peaking at 1.12\% in the 40\%--60\% range, before declining at higher levels: 0.81\% at 60\%--80\%, 0.71\% at 80\%--99\%, and reaching its lowest rate of 0.21\% when project contributors made 100\% of the comments. This inverted-U pattern suggests that discussions with moderate contributor presence, i.e., where external and internal participants interact most, may be particularly prone to friction, while discussions dominated by project contributors are associated with substantially less toxicity.

\begin{table}[tb]
    \small
    \caption{Toxic conversation rates across discussions with varying levels of project contributor participation.}
    \centering
    \begin{tabular}{l|c}
        \hline
        \textbf{Comments by} & \textbf{Toxic} \\
        \textbf{Project Contributors (\%)} & \textbf{Conversations (\%)} \\
        \hline
        0\%--20\%    & 0.95\% (203/21339) \\
        20\%--40\%   & 1.02\% (194/19002)\\
        40\%--60\%   & 1.12\% (370/32911)\\
        60\%--80\%   & 0.81\% (79/9712)\\
        80\%--99\%  & 0.71\% (16/2256)\\
        100\%        & 0.21\% (84/39537)\\
        \hline
    \end{tabular}
    \label{tab:developer_toxicity}
\end{table}

% \vspace{-1mm}
\begin{observecomment}{}
{
Our findings partially support Miller et al. and Imran et al.’s observation that greater project contributor engagement correlates with lower toxicity. External participants made 44.48\% of all comments, but 57.56\% in toxic conversations. However, the relationship is non-monotonic: toxicity peaks at moderate contributor participation (40\%--60\%) before declining at higher levels, suggesting that mixed-participation discussions may introduce friction, while predominantly contributor-driven discussions are the most civil.
}
\end{observecomment}

\subsection{Toxicity and Mention Graph}
Saveski et al.~\cite{saveski_structure_2021} examined the relationship between conversation structure, specifically reply trees, and toxicity in discussions on X (formerly Twitter). They represented conversations as reply trees, where the root node was the initial tweet that initiated the discussion, and the subsequent nodes were replies. They hypothesized that larger reply trees tend to be more toxic and found a positive correlation between tree size and toxicity. They measured toxicity as the fraction of toxic tweets within a conversation and analyzed how toxicity levels varied with conversation size, defined as the total number of tweets. Their findings suggested that as reply structures expanded, discussions became increasingly toxic.

% \begin{figure}[t] % 'H' forces the figure to appear exactly here (requires 'float' package)
%     \centering
%     \includegraphics[width=1\linewidth]{Figures/toxicity_by_path_length.png} % Adjust the width as needed
%     \caption{Analysis of toxicity patterns in GitHub issues by mention graph path depth. (a) Percentage of issues that are toxic within each mention path depth category, revealing how toxicity rates vary across different path lengths. (b) Distribution of all toxic issues across mention path depths, showing what percentage of the overall toxic issues fall into each path length category.}
%     \label{fig:toxicity_by_path_length}
% \end{figure}

GitHub issue and PR discussions do not support threaded replies in the way that X does. To evaluate whether a similar relationship between conversation depth and toxicity exists on GitHub, we constructed mention graphs for each conversation in our dataset. In our approach, each unique commenter is represented as a node, with a directed edge from commenter A to commenter B if A explicitly mentions B using the “@B” syntax in a comment. The longest path length in a mention graph is defined as the deepest chain of mentions within a conversation. Unlike reply trees, which capture hierarchical reply structures based on direct responses, mention graphs represent explicit participant references within discussions, independent of the reply structure. While most toxic conversations appear in discussions with shallow mention depth, likely because they are much more common overall, the rate of toxicity increases with greater depth. 
As shown in Table~\ref{tab:mention_graph_toxicity}, toxicity increases with mention depth: it remains below 1.05\% for path lengths 0–2, rises to 2.46\% at depth 3, and further increases to 3.11\% for deeper chains ($\geq 4$).

\begin{observecomment}{}
{
Our findings support the general observation that deeper mention chains tend to be more toxic. Although such deep chains are relatively rare, conversations with longer mention paths ($\geq 4$) exhibit higher toxicity rates (3.11\%) compared to shallow discussions. This suggests that increasingly complex interaction patterns are more prone to toxic behavior, consistent with prior research.
}
\end{observecomment}

\subsection{Prevalence and Trends of Toxic Conversations Over Time}
Previous work by Raman et al. found that toxic conversations are rare in GitHub discussions, occurring in about 0.6\% of cases~\cite{raman2020stress}. 
They conducted a longitudinal study of 1,732,124 GitHub conversations collected from 2012 to 2018. They initially manually annotated a small subset of 611 issues (of which 167 were toxic threads) and trained a machine learning classifier to detect toxicity in GitHub. 
After applying the classifier to the larger dataset, they concluded that toxic conversations are infrequent. Their analysis also found that toxicity had a declining trend over time.

In contrast, we use a HITL-constructed dataset drawn from a more recent corpus of GitHub issue and PR discussions. In our dataset, \totaltoxic\ conversations (0.76\%) were toxic, slightly higher than the 0.6\% reported by Raman et al. However, this comparison should be interpreted with caution: the two studies use different detection methodologies (HITL with LLM annotation vs.\ a trained ML classifier), different sampling strategies, and different repository pools, so the difference in rates may reflect methodological variation rather than a true temporal shift. That said, a recent 2024 GitHub survey offers suggestive---though not directly comparable---corroboration, noting that 64.23\% of developers either experienced or witnessed negative interactions, up from 60\% in 2017~\cite{GitHub_GitHub_Open_Source_2024, GitHubOpenSourceSurvey2017}. Together, these data points leave open the possibility that toxicity has not declined as Raman et al.\ observed in their earlier period, though stronger longitudinal evidence is needed to draw firm conclusions.

\begin{table}[tb]
    \small
    \caption{Toxic conversations by depth of mention chains.}  
    \centering
    \begin{tabular}{c l}
        \hline
        \textbf{Deepest Path Length} & \textbf{Toxic Conversations (\%)} \\
        \hline
        0  & 0.71\% (611/86179) \\
        1  & 0.77\% (241/31478) \\
        2  & 1.05\% (61/5827) \\
        3  & 2.46\% (25/1016) \\
        % 4  & 2.78\% (5/180) \\
        $\geq 4$  & 3.11\% (8/257) \\
        % $\geq 5$ & 3.90\% (3/77) \\
        \hline
    \end{tabular}  
    \label{tab:mention_graph_toxicity}
\end{table}

\begin{figure}[tb] 
    \centering
    \includegraphics[width=0.9\linewidth]{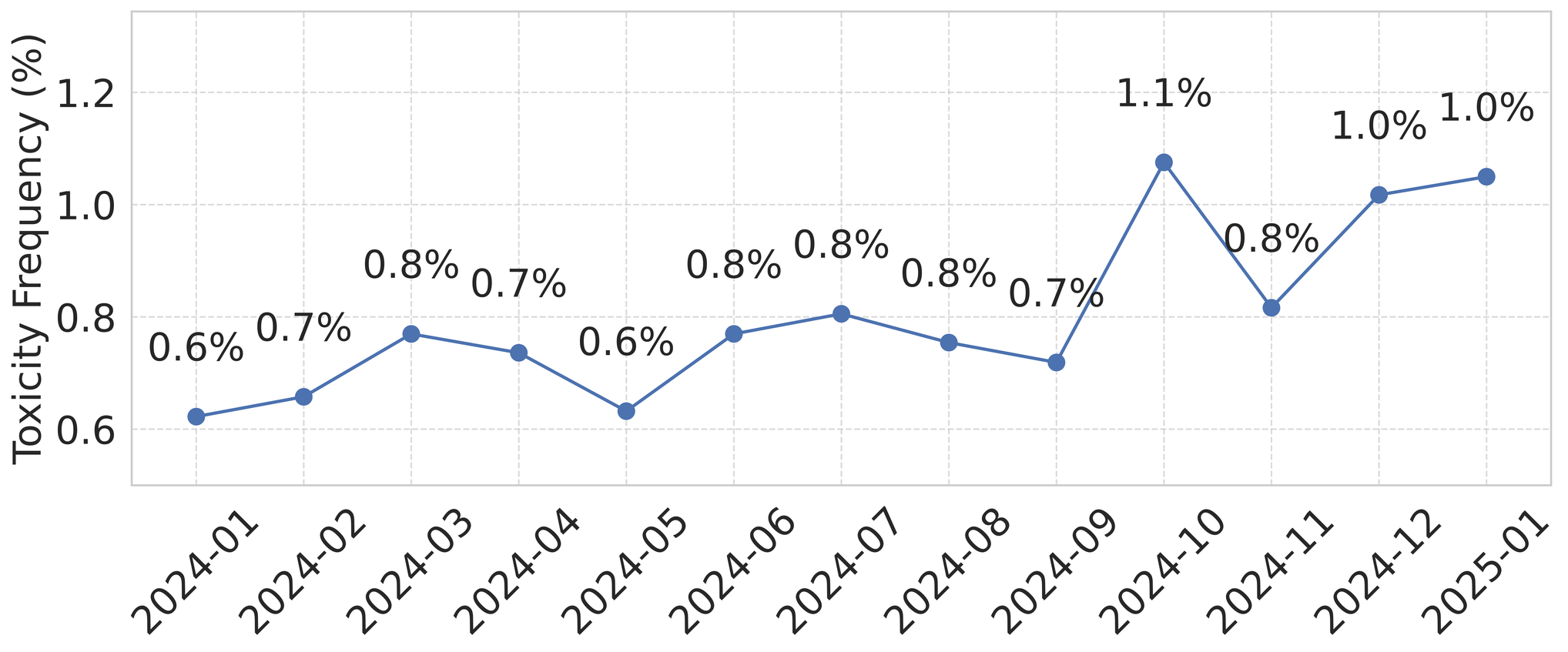} % Adjust the width as needed
    \caption{Monthly percentage of toxic conversations in GitHub discussions (Jan.\ 2024 -- Jan.\ 2025).}
    \label{fig:toxicity_trend_2024}
\end{figure}

To further examine recent trends in toxicity, we conducted a month-by-month analysis of GitHub conversations from January 2024 through January 2025. In Figure~\ref{fig:toxicity_trend_2024}, we show the monthly toxicity frequency measurements, calculated as the proportion of toxic conversations each month.
As shown in Figure~\ref{fig:toxicity_trend_2024}, toxicity remains consistently low but exhibits mild fluctuations over time. The toxicity rate ranges from approximately 0.62\% to 1.08\%, with most months clustered between 0.7\% and 0.8\%. Notably, a slight increase is observed toward the end of 2024, with peaks around October (1.08\%), December (1.02\%), and January 2025 (1.05\%), suggesting a modest upward trend rather than complete stability.

\begin{observecomment}{}
{
Our findings remain consistent with Raman et al., who reported that toxic conversations are relatively rare in GitHub discussions~\cite{raman2020stress}. Our observed average toxicity rate of approximately 0.76\% is higher than their reported 0.6\%, though as noted above, this difference may partly reflect methodological variation rather than a true temporal shift. The monthly trend indicates mild temporal variation, with a small upward shift toward the end of the observed period.
}
\end{observecomment}

\subsection{Relation Between Toxicity and Number of Comments}

Previous work by Imran et al.~\cite{imran2026toxicity} found that toxic conversations on GitHub tend to be longer, with a median of 11 comments, compared to 6 in non-toxic discussions (out of 202 toxic and 696 non-toxic threads). 
They found that toxicity often appeared within the first 10 comments and that over half of toxic discussions contained multiple toxic comments. This suggests that toxicity arises more often in longer discussions and contributes to prolonging them.

To assess whether this trend holds in our dataset, we analyzed {\totalCount} GitHub issue/PR conversations, comparing the {\totaltoxic} toxic ones against 123,811 non-toxic ones. At scale, the association between toxicity and conversation length is considerably weaker than previously reported. The median number of comments in toxic conversations was 4, compared to 3 in non-toxic ones---a gap of just one comment, far smaller than the 11 vs.\ 6 reported by Imran et al. The distributional difference is somewhat more visible at the 75th percentile (8 comments for toxic vs.\ 5 for non-toxic), and toxic conversations exhibited greater variability in length (SD of 7.64 vs.\ 4). Still, the overall effect is modest, suggesting that while toxicity and conversation length are associated, the relationship is weaker than small-scale studies indicated.

\begin{observecomment}{}
{
At scale, the association between toxicity and conversation length is much weaker than Imran et al.\ reported (median of 4 vs.\ 3, compared to their 11 vs.\ 6). The directional trend holds, but it suggests that conversation length only modestly correlates with toxicity rather than strongly.
}
\end{observecomment}

\subsection{Pull Request Acceptance and Toxicity in Discussions}
Previous work by Sarker et al. examined the relationship between pull request (PR) acceptance and toxicity in OSS development~\cite{sarker2025landscape}. They analyzed a sampled subset from a collection of 6.3 million PRs from 1,155 projects. Their study found that accepted PRs are less likely to encounter toxicity, while rejected PRs tend to have a higher likelihood of toxic interactions. Specifically, they found that 2.47\% of accepted PRs contained at least one toxic comment, compared to 3.26\% of rejected PRs.

To evaluate whether a similar relationship holds in our dataset, we analyzed 39,891 PR conversations and measured toxicity across accepted and non-accepted PRs. Our findings align with the hypothesis that accepted PRs are less toxic. Of the 24,656 accepted PRs, 57 (0.23\%) contained toxic comments, whereas 155 (1.19\%) of the 12,999 non-accepted PRs were toxic -- over five times higher. 

From a toxic PR perspective, we found that 25.56\% of all toxic PRs were accepted, whereas 62.01\% of all non-toxic PRs were accepted. This reinforces the idea that toxic interactions are more prevalent in rejected PRs, while accepted PRs experience low toxicity.

\begin{observecomment}{}
{
Our findings support the hypothesis previously observed that accepted PRs are less likely to encounter toxicity. In our corpus, toxicity rates in non-accepted PRs (1.19\%) were higher than accepted PRs (0.23\%). Additionally, toxic PRs had a lower acceptance rate (25.56\%) compared to non-toxic ones (62.01\%).
}
\end{observecomment}

\section{Novel Findings on Toxicity Dynamics}

Our large-scale dataset provides new opportunities to study the relationship between toxicity and project and contributor dynamics on GitHub. We describe each finding below.

\subsection{Programming Language and Toxicity Rates}

While programming languages are not themselves the source of toxicity, they often correlate with broader community norms and dynamics. Different programming language ecosystems attract distinct profiles, project domains, and cultural norms~\cite{annunziata2024empirical, storey2016social}.

\begin{table}[tb]
\small
\caption{Toxicity rates by programming language in languages with $\geq$10,000 issues and $\geq$20 projects.}
\centering
\begin{tabular}{l c c c c}
\hline
\textbf{Primary} & \textbf{Repo} & \textbf{Issues} & \textbf{Toxic Issues} & \textbf{Toxicity} \\
\textbf{Language} & \textbf{Count} & \textbf{Count} & \textbf{Count} & \textbf{Ratio (\%)} \\
\hline
Python      & \added{34} & \added{12,407} & \added{132} & \added{1.06} \\
C++         & \added{20} & \added{15,329} & \added{125} & \added{0.82} \\
Go          & \added{30} & \added{12,189} & \added{93}  & \added{0.76} \\
JavaScript  & \added{37} & \added{13,852} & \added{90}  & \added{0.65} \\
TypeScript  & \added{74} & \added{28,862} & \added{166} & \added{0.58} \\
\hline
\end{tabular}
\label{tab:toxicity_by_language}
\end{table}

We examined toxicity rates in repositories grouped by their primary programming language as provided through GitHub REST API. We limited our analysis to languages represented by at least 20 repositories, each with a minimum of 10,000 issues and PRs, to ensure representativeness and statistical reliability. Table~\ref{tab:toxicity_by_language} summarizes the results. Python and C++ repositories exhibit comparatively higher toxicity rates (\added{1.06\%} and \added{0.82\%}, respectively). Go and JavaScript show moderate levels of toxicity (\added{0.76\%} and \added{0.65\%}), while TypeScript has the lowest observed toxicity rate at \added{0.58\%}.

\added{To assess whether the observed differences across ecosystems are statistically reliable, we applied a Kruskal-Wallis test on per-repository toxicity rates within each language group. The test yielded $H = 9.004$, $p = 0.061$, marginally non-significant at $\alpha = 0.05$. The ordering of ecosystems (Python $>$ C++ $>$ Go $>$ JavaScript $>$ TypeScript) remains consistent with the aggregate rates in Table~\ref{tab:toxicity_by_language}.}

\begin{observecomment}{}
{
      Toxicity levels vary across programming language ecosystems. Python and C++ repositories show relatively higher toxicity rates, Go and JavaScript fall in a moderate range, and TypeScript exhibits the lowest toxicity. \added{However, the differences do not reach statistical significance ($p = 0.061$).}
}
\end{observecomment}

\subsection{Toxicity and Project Governance}\label{sec:governance}

Community smells in open source software projects often reflect structural or social dysfunctions that hinder collaboration and inclusivity. Effects such as the ``Prima Donna'' and ``Lone Wolf'' are commonly associated with weak or poorly defined governance structures~\cite{tamburri2019exploring, almarimi2020detection, tamburri2015social, annunziata2024empirical}. Prior work also suggests that stronger governance, including active moderation~\cite{hsieh2023nip}, codes of conduct~\cite{li2021code}, and organizational support~\cite{sarker2025landscape}, can reduce harmful interactions. Motivated by this literature, we examine how governance-related characteristics align with toxicity in our dataset.

\begin{figure}
\centering
\includegraphics[width=1.0\linewidth]{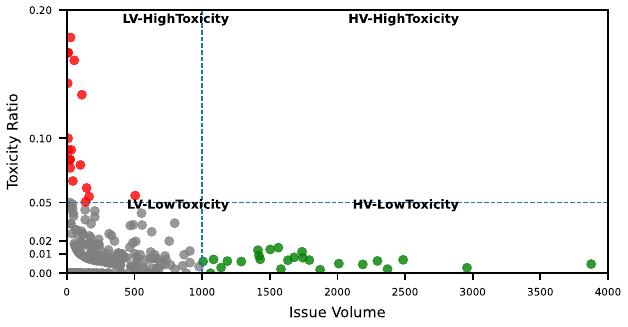}
\caption{Repository groups by issue activity and proportion of toxic discussions.}
\label{fig:toxic-clusters}
\end{figure}

We partition repositories along two interpretable dimensions: new issue volume during the study period and the proportion of toxic issues. We set fixed thresholds on these axes by manually inspecting the empirical distribution rather than applying learned clustering or automatic optimization, assigning repositories to four quadrants. This rule-based grouping preserves interpretability and makes categories easy to examine substantively. We retain all repositories, since highly toxic cases are central to the analysis rather than measurement noise. Figure~\ref{fig:toxic-clusters} shows the full dataset partitioned by issue volume and toxicity ratio.

Figure~\ref{fig:toxic-clusters} reveals three populated quadrants. The largest, Low Volume-Low Toxicity (LV-LT), contains 292 repositories with modest activity and low toxicity. High Volume-Low Toxicity (HV-LT) contains 27 repositories with high issue activity but still low toxicity. Low Volume-High Toxicity (LV-HT) contains 21 repositories with lower activity but substantially higher toxicity. No repositories fall into the High Volume-High Toxicity quadrant, suggesting that sustained toxicity is uncommon among highly active repositories in our sample.

To better understand elevated toxicity in LV-HT, we examine governance through administrator participation in toxic discussions. Since maintainer-level data are unavailable via the GitHub REST API, we proxy administrators using users with write or higher access (Owners, Members, and Collaborators), excluding Contributors, who lack write, push, or merge permissions.

Clear differences emerge across groups. HV-LT repositories have more administrators participation, with a median of 23, 
compared with 4 in LV-LT and 1 in LV-HT. This suggest that LV-HT repositories are typically managed by smaller and less active core teams
We also examine repository Health Percentage, which reflects the presence of elements such as a code of conduct, contributing guidelines, a license, and issue and pull request templates. LV-HT has the lowest median health percentage at 75.0\%, compared with 87.0\% for HV-LT and 85.0\% for LV-LT.

Ownership structure provides further context. In HV-LT, 100\% of repositories are organization-owned. In LV-LT, 87.4\% are organization-owned and 12.6\% are user-owned. LV-HT has the lowest share of organization-owned repositories at 66.7\%, with 33.3\% owned by individual users.

Although we do not directly measure governance, these groups appear to reflect distinct governance styles. HV-LT repositories resemble structured, team-based governance models, such as Apache-style projects~\cite{mockus2002two}. In contrast, LV-HT repositories align more closely with centralized models led by a small core of maintainers~\cite{fogel2005producing}. LV-LT may reflect lightweight or emerging governance structures~\cite{o2007emergence}.

\begin{observecomment}{}
{
Repositories with less structured governance tend to exhibit higher toxicity. In the Low Issue Volume--High Toxicity (LV-HT) group, the median number of admins was \added{1} (vs. 23 in HV-LT) and the median GitHub Health Score was \added{75.0}\% (vs. \added{87.0}\%). Together, this suggests that GitHub projects with more structured governance and larger teams tend to experience lower levels of conversational toxicity.
}
\end{observecomment}

\subsection{Behavior of Highly Active GitHub Users}

To understand the behavior of highly active GitHub users in toxic discussions, we analyzed individuals who had participated in at least 10 toxic GitHub issues in our dataset. There were 16 such users, each appearing in 10–24 toxic conversations. We examined their repository affiliations, administrative roles (i.e., users with write or higher repository access), and dual roles as both participants and moderators in contentious discussions.

All 16 were either repository administrators (15/16) or contributors (1/16) to at least one repository, and most were core project members. On average, they participated in 10,005 issues (median of 388), with the most active user involved in 3,734 issues. Together, they appeared in 236 toxic threads (24.95\%, 236/946) and authored 540 comments across those conversations. Two authors independently annotated these comments to assess: (i) whether users authored toxic comments; and (ii) whether they attempted moderation in response to toxicity.

The most common pattern was the overlap between writing toxic comments and issuing moderation responses. All 15 administrators did both, often with moderation that was itself toxic in tone. This challenges the assumption that administrators take "appropriate and fair corrective action"~\cite{roundtree2022facial, contributorcovenantv1.4}, suggesting instead that authority figures engage in the same behaviors they are expected to moderate. The one non-administrator also authored toxic comments.

Qualitative analysis showed that toxic and moderating behaviors were often intertwined. Administrators sometimes responded to user queries with hostility or sarcasm framed as moderation. For example, one maintainer replied to issue spam with, \textit{"Stop doing this. You are repeating creating this. If you keep doing this, I have to ask the [NAME] to ban you. Last warning!"} Another closed a pull request stating, \textit{"Closing this PR as I don’t enjoy being a guinea pig,"} citing template violations. In another case, a moderator dismissed a user’s parser concerns with, \textit{"Have you ever written a parser of anything in your life? [...] That alone tells me all I need to know about your level of competence. Good luck to you."} These examples show that moderation can take a punitive, condescending, or dismissive tone. In some cases, it escalated hostility rather than reducing it. One particularly aggressive comment stated, \textit{"Shut up. Go install Windows 10 instead of tirelessly begging for features... Get a life."} Another dismissive comment read, \textit{``[...] Nothing much of rocket science here. [...]"} Language like this, especially when paired with actions and threats such as locking conversations or blocking users, does not reflect the communication standards typically expected in responsible OSS governance.

% Notably, these behaviors spanned high-profile projects with thousands of stars (median X, following~\cite{dabbish2012social}). At the individual level, the 16 users had between X and X followers (median X), indicating that toxicity and punitive moderation can occur at the highest levels of project leadership.

\begin{observecomment}{}
{
A focused analysis of 16 highly active GitHub users (15 administrators, 1 contributor) shows that all authored both toxic and moderation comments. This challenges the view of administrators as fair corrective enforcers and suggests leadership can itself be a source of toxicity.
}
\end{observecomment}
\section{Effect Sizes and the Impact of Manual Review}\label{sec:effect_sizes}
\added{To assess the downstream impact of manual review, we computed effect sizes for five quantitative estimates derived from Observations~\#1 and~\#3--\#5 under two label sets, the manually corrected labels (946 toxic conversations) and the LLM-only labels (805 toxic conversations). Manual review changed 169 labels in total, relabeling 155 conversations from non-toxic to toxic and 14 from toxic to non-toxic, for a net increase of 141 toxic conversations. We omit the other observations from this analysis for various reasons. For Observation~\#2, the toxicity rates by mention depth in Table~\ref{tab:mention_graph_toxicity} already convey the size of the effect. Observations~\#6 and~\#7 are assessed at the repository level in Section~5, and Observation~\#8 describes a qualitative pattern among named individuals.}

\added{We use Cliff's~$\delta$ for two-group comparisons, Spearman~$\rho$ for rank-based correlations, and Cram\'er's~$V$ for categorical associations. Cliff's~$\delta$ ranges from $-1$ to $+1$ and expresses how often values in toxic conversations exceed those in non-toxic ones. Positive values indicate that toxic conversations score higher on the measure, and $V$ measures association strength on a 0--1 scale. Following conventional thresholds, $|\delta|$ values of roughly $0.15$, $0.33$, and $0.47$ mark the onset of small, medium, and large effects. For $\rho$ and $V$ the corresponding values are $0.1$, $0.3$, and $0.5$. We also include the overall toxicity prevalence (\#3a), a descriptive proportion rather than an association measure. The trend estimate (\#3b) is computed over monthly rates and all other estimates over conversations. Table~\ref{tab:effect_sizes} reports each effect size under the corrected labels, together with its change relative to the LLM-only labels, computed as corrected minus LLM-only.}

\added{Most effects are small, as expected given that toxicity is rare and no single factor drives it. The strongest conversation-level association is the participant-type finding (Observation~\#1, Cliff's~$\delta = -0.232$, a small effect). Among non-tied random toxic/non-toxic pairs, the non-toxic conversation has higher developer participation $\sim$62\% of the time. The toxicity trend (Observation~\#3b, Spearman~$\rho = +0.775$) is large by these thresholds. Rank correlation captures how consistently the monthly rates rise rather than how steeply, so this estimate is compatible with the modest upward shift described in Section~4.3. It also rests on only 13 monthly data points and should be interpreted with caution. For PR acceptance (\#5), non-accepted PRs are 5 times more likely to be toxic, yet Cram\'er's~$V = 0.061$ falls below the small-effect threshold. This occurs because toxicity remains rare in both groups, so a large relative risk corresponds to a small absolute association.}

\added{Manual review materially changed two estimates. For the prevalence estimate (\#3a), both label sets are scored on the same 124{,}757 conversations, so McNemar's test is the appropriate comparison. The corrections were strongly asymmetric, with 155 conversations gained and 14 lost, yielding a matched-pairs odds ratio of $11.07$ (95\% CI (confidence interval) of $[6.41, 19.13]$, $p < 0.001$). The interval lies entirely above 1, indicating a systematic shift. For the conversation-length estimate (\#4), the effect is $2.5\times$ larger after manual review (Cliff's~$\delta$ rises from $0.077$ to $0.195$), crossing from below to above the small-effect threshold. The reason is that the conversations the LLM missed came disproportionately from longer discussions. Without manual review, we would have reported a substantially weaker association between conversation length and toxicity. The remaining three estimates shift only slightly under relabeling, so their interpretations are unchanged.}
\begin{table}[tb]
\small
\caption{\added{Effect sizes for five quantitative estimates
derived from Observations~\#1 and~\#3--\#5,
computed on the manually corrected labels (946 toxic conversations).
Change is the corrected-label estimate minus the LLM-only estimate
(805 toxic conversations); pp~=~percentage points.}}
\centering
\resizebox{\columnwidth}{!}{%
\begin{tabular}{l l l r r}
\hline
\textbf{Obs.} & \textbf{Finding} & \textbf{Metric} & \textbf{Effect Size} & \textbf{Change} \\
\hline
\#1  & Developer engagement   & Cliff's $\delta$  & $-0.232$ & $+0.020$ \\
\rowcolor{gray!20}
\#3a & Toxicity prevalence    & Proportion        & $0.76\%$ & $+0.11$\,pp \\
\#3b & Toxicity trend         & Spearman $\rho$   & $+0.775$ & $-0.050$ \\
\rowcolor{gray!20}
\#4  & Conversation length    & Cliff's $\delta$  & $+0.195$ & $+0.118$ \\
\#5  & PR acceptance          & Cram\'er's $V$    & $0.061$  & $+0.004$ \\
\hline
\end{tabular}}
\label{tab:effect_sizes}
\end{table}
\section{Related Work}

Our work builds upon prior work in two main areas: toxicity analysis in OSS development and HITL approaches in data annotation using a LLM. Below we discuss these two key areas.

\subsection{Toxicity in OSS}

Toxicity-free OSS has been a long-standing goal of the community~\cite{carillo2016towards, filippova2015mudslinging, filippova2016effects}. The study of toxicity and negative interactions in OSS ecosystems has received significant attention in recent years.
Researchers have explored various forms of negative engagement across multiple communication channels, including issue trackers, pull requests, chat discussions, and code reviews. 
Early research predominantly focused on sentiment analysis in OSS communication, identifying positive and negative expressions. Over time, the focus has expanded towards more nuanced aspects, such as granular-level emotions~\cite{imran2022data, imran_2024_emocause, novielli2014towards, ortu2015bullies}, incivility~\cite{ferreira2021shut, rahman2024words, ehsani2024incivility, ferreira2022how, ferreira2024incivility}, and toxicity~\cite{sarker_automated_2023, sarker2022identification, sarker2025landscape, ehsani2024analyzing, raman2020stress, jamieson2024predicting, miller_did_2022}. Prior work also shows that LLMs can miss subtle emotional dynamics in software engineering contexts, such as in pair programming conversations~\cite{rahman2026pair}.

Researchers have continued to explore this area further and investigate even more nuanced aspects.
Li et al. found that codes of conduct are used to govern offensive forms of speech~\cite{li2021code}. Qiu et al. noted interpersonal conflict in code reviews~\cite{qiu2022detecting}. Qiu et al. later proposed a dashboard that overviews community health, that includes toxicity score of the comments~\cite{qiu_climate_2023}. Trinkenreich et al. conducted a study on understanding the link between the motivation of OSS developers and the sense of belonging to the community~\cite{trinkenreich2023belong}.
Rahman et al.'s survey of 171 developers revealed that half of the developers (56\%) have encountered workplace uncivil discourse and the majority of this group (83.70\%) experiences it at least once a month~\cite{rahman2024words}. Gunawardena et al. observed that destructive code reviews have a negative impact~\cite{gunawardena2022destructive}. Ferreira et al. noted inappropriate feedback in the Linux kernel mailing list~\cite{ferreira2021shut}. Imran et al. experimented on proactive toxicity detection in GitHub conversations~\cite{imran2026toxicity}. 
In a 2020 study, Raman et al. conducted a longitudinal study on finding toxicity trends~\cite{raman2020stress}. In a recent study, Sarker et al. conducted a large-scale study on GitHub toxicity using their ToxiCR tool as a toxicity identifier~\cite{sarker2025landscape}. Unlike those studies, we proposed a novel human-in-the-loop approach to annotate toxicity in GitHub and attempted to scale up previous observations on toxicity in OSS. 

\subsection{Human-in-the-Loop Data Annotation Strategies}

The human-in-the-loop paradigm has emerged as a crucial methodology for improving the quality and reliability of data annotation for machine learning tasks. Several studies have demonstrated the effectiveness of combining human expertise with automated systems to create more accurate and contextually appropriate labeled datasets~\cite{holzinger2016interactive, monarch2021human, wang2021putting, wu2022survey, zhang2023llmaaa, wang2024human}. With the advancement of LLMs in recent years, the integration of LLMs into HITL pipelines has been widely attempted~\cite{wang2024human, artemova2024hands, wang2021putting, wu2022survey, pangakis2025keeping, kim2024meganno+}. Studies show that ChatGPT can outperform crowd workers and significantly reduce manual effort~\cite{gilardi2023chatgpt}. Some studies have found that integrating LLMs as preliminary annotators in HITL pipelines can increase annotation throughput while maintaining comparable quality to fully manual approaches~\cite{karpinska2023large, he2024annollm}. Unlike previous studies, ours is one of the first large-scale toxicity analyses on GitHub that leverages generative LLMs for toxicity annotation to conduct an empirical study. 
\section{Limitations}

\noindent
{\em Construct Validity.}
Our reliance on LLM-generated toxicity annotations may introduce biases or misclassifications despite human verification; the HITL validation model mitigates this by routing likely errors to manual review. Our definition of toxicity follows prior research but may not capture all harmful interactions; we address this by incorporating a range of toxicity categories grounded in established literature. Finally, our approach considers only textual content, ignoring images, code snippets, or external links that might provide additional context. \added{Toxicity annotation also involves inherent subjectivity; we mitigate this through a guideline-based protocol grounded in prior GitHub toxicity studies, annotators with domain expertise, and a Cohen's $\kappa$ of 0.78 on the shared annotation pilot, indicating substantial agreement.}

\noindent
{\em Internal Validity.}
The validation model’s performance is constrained by the quality and representativeness of its training data (898 conversations); we mitigate this by evaluating across multiple classification thresholds and on an independent dataset. \added{The training data also predates our 2024 corpus, raising a potential distribution-shift concern; we address this through two complementary checks: external validation on Raman et al.’s independently sampled dataset (weighted F1 = 0.96, Section~\ref{sec:annot_model}) and a manual audit of 5,000 unflagged conversations from the 2024 corpus itself, which found only 11 errors (overall labeling accuracy 99.90\%, Section~3.3).} Our repository selection criteria (star count, activity level) may introduce selection bias, potentially overlooking toxicity patterns in smaller or less active projects.

\noindent
{\em External Validity.}
Our findings may not generalize beyond GitHub to platforms such as GitLab or Bitbucket, though we select repositories across diverse domains and programming languages to ensure broad coverage within GitHub. The dataset’s time frame (January 2024 to January 2025) provides a snapshot rather than a longitudinal view; future work could assess temporal changes over longer periods. Differences in annotation methodologies between our study and prior work may also affect direct comparisons; further replication is needed to confirm the stability of these trends.

\section{Conclusions and Future Work}
Our large-scale investigation offers new insights into toxicity in GitHub discussions. Using a human-in-the-loop annotation pipeline, we analyzed over {\totalCount} conversations from 340 repositories, producing one of the most comprehensive toxicity datasets in OSS research to date.

We found that toxicity is slightly more prevalent (0.76\%) than previously reported (0.6\%). Our analysis validates several prior findings at scale: toxicity peaks at moderate contributor participation before declining at higher levels, increases in deeply threaded conversations, and is more common in longer discussions and rejected pull requests. We also highlight novel findings about the relationship between toxicity and project dynamics. We observed variation in toxicity across language ecosystems, with Python and C++ showing higher rates than TypeScript or JavaScript, though these differences did not reach statistical significance. Repositories with weaker governance, i.e., fewer admins, limited organizational backing, and lower health scores, exhibited substantially higher toxicity. 

Future work will extend our study longitudinally to capture evolving patterns in toxicity, compare community-specific norms across GitHub ecosystems, and develop adaptive intervention mechanisms based on context and contributor history. We also plan to investigate how toxicity affects contributor retention and project sustainability.

\section*{Data Availability Statement}
Our replication package is available on Zenodo at
\color{blue}\url{https://doi.org/10.5281/zenodo.21231251}\color{black}.

\balance

\bibliographystyle{ACM-Reference-Format}
\bibliography{references,toxicity}

\end{document}